\documentclass[aip,jcp,amsmath,amssymb,onecolumn,groupedaddress]{revtex4-2}
\usepackage{graphicx}
\usepackage{amsmath}
\usepackage{graphics}
\usepackage{colortbl}
\usepackage{xcolor}
\usepackage{tabularx}
\usepackage{dsfont}
\usepackage{natbib}

\usepackage{amsmath}
\usepackage{amsfonts}
\usepackage{amssymb}
\usepackage{xcolor}
\usepackage{subcaption}
\DeclareUnicodeCharacter{2212}{-}

\usepackage[normalem]{ulem}

\usepackage[bookmarks=true]{hyperref}
\hypersetup{colorlinks=true, citecolor=blue, urlcolor=blue, linkcolor=blue}

\newcommand*{\citen}[1]{%
  \begingroup
    \romannumeral-`\x 
    \setcitestyle{numbers}%
    [\cite{#1}]%
  \endgroup   
}

\begin{document}

\title{Nonadiabatic Quantum Dynamics of Molecules Scattering from Metal Surfaces}

\author{{Riley J. Preston$^{1}$}$^{*}$, Yaling Ke$^2$, Samuel L. Rudge$^1$, Nils Hertl$^{3,4}$, Raffaele Borrelli$^5$, Reinhard J. Maurer$^{3,4}$, Michael Thoss$^1$}

\affiliation{$^1$Institute of Physics, University of Freiburg, Hermann-Herder-Strasse 3, 79104 Freiburg, Germany\\
$^2$Department of Chemistry and Applied Biosciences, ETH Z\"urich, 8093 Z\"urich, Switzerland\\
$^3$Department of Chemistry, University of Warwick, Gibbet Hill Road, CV4 7AL, Coventry, United Kingdom\\
$^4$Department of Physics, University of Warwick, Gibbet Hill Road, CV4 7AL, Coventry, United Kingdom\\
$^5$DISAFA, University of Torino, I-10095 Grugliasco, Italy}

\begin{abstract}
\noindent Nonadiabatic coupling between electrons and molecular motion at metal surfaces leads to energy dissipation and dynamical steering effects during chemical surface dynamics. We present a theoretical approach to the scattering of molecules from metal surfaces that incorporates \emph{all} nonadiabatic and quantum nuclear effects due to the coupling of the molecular degrees of freedom to the electrons in the metal. This is achieved with the hierarchical equations of motion (HEOM) approach combined with a matrix product state representation in twin space. The method is applied to the scattering of nitric oxide from Au(111), for which strongly nonadiabatic energy loss during scattering has been experimentally observed, thus presenting a significant theoretical challenge. Since the HEOM approach treats the molecule-surface coupling exactly, it captures the interplay between nonadiabatic and quantum nuclear effects. Finally, the data obtained by the HEOM approach is used as a rigorous benchmark to assess various mixed quantum-classical methods, from which we derive insights into the mechanisms of energy dissipation and the suitable working regimes of each method.
\end{abstract}

\maketitle

\newpage
\section{Introduction}

\noindent Understanding the dynamics of molecules interacting with metal surfaces is vital in a wide range of scenarios, such as reactive and catalytic processes \cite{saalfrank06,brandbyge95,kim02}, charge transport through molecular nanojunctions \cite{preston23,preston2020,erpenbeck23,yaling23}, and scattering experiments \cite{bunermann2015,kroes16,tully2000,rahinov2024}. Often critical in these setups is the influence of the metal electrons on the dynamics of the molecule, in particular, due to nonadiabatic processes that lead to the creation of electron-hole pair (EHP) excitations within the metal. For example, nonadiabatic processes lead to vibrationally inelastic scattering \cite{huang2000}, chemically-induced currents \cite{gergen2001}, dynamical steering effects \cite{shenvi09}, and ultrashort vibrational lifetimes \cite{chang1990,morin1992,lass2005}. Consequently, theoretical treatments of such situations require sophisticated methods capable of going beyond the Born-Oppenheimer approximation \cite{wodtke06}.

The scattering of molecules from metal surfaces is a prototypical example of surface dynamics and has motivated the development of a plethora of nonadiabatic theoretical methods \cite{kroes16}. These range from mixed quantum-classical (MQC) approaches that treat the motion of atomic nuclei classically  while being coupled to quantum electronic degrees of freedom, to fully quantum approaches. To the first category belong methods such as Ehrenfest dynamics \cite{grotemeyer14,tully23,preston22}, electronic friction approaches \cite{spiering19,maurer19,box21,dou18,loncaric17,fuchsel13}, and surface-hopping techniques \cite{miao19,shenvi12,gardner23_2}. Fully quantum treatments, in contrast, are able to propagate the many-dimensional nuclear wavepacket, but generally rely on an approximate treatment of the coupling to surface electrons \cite{serwatka20,nest2000,monturet10} or neglect the role of electronic excitations in the metal entirely \cite{saalfrank93,jiang14_2,crespos06}.

While these methods have had impressive success for a multitude of scattering problems, they often fail to describe the full complexity of the dynamics. For instance, one of the most well-explored scattering experiments for probing nonadiabatic effects is that of an NO molecule scattering from Au(111). Here, it has been shown that the coupling to EHPs in the surface strongly impacts the dynamics of the scattering molecule \cite{golibrzuch13,wodtke16,Golibrzuch14}, leading to multi-quantum relaxation in the vibrational state of the molecular bond \cite{kruger15}. However, theoretical treatments of NO scattering are often unable to accurately capture the dynamics observed in experiments, usually underpredicting the amount of vibrational relaxation \cite{kruger15}. The origins of this failure are hard to diagnose due to the combination of different sources of errors in the description of the electronic structure or the coupled electron-nuclear dynamics of realistic systems.  This problem can only be tackled by separately benchmarking the quality of electronic structure predictions~\cite{gerrits_density_2020} and by quantifying the basic ability of the employed nonadiabatic dynamics method to capture all relevant dynamic effects. Unfortunately, existing MQC methods vary widely in their quantitative predictions\cite{gardner23}.

This is the primary motivation for this work, in which we employ a fully quantum mechanical approach for simulating the scattering of molecules from metal surfaces that incorporates the molecule-surface interaction exactly. This is achieved via the hierarchical equations of motion (HEOM) method, which is a numerically exact technique for modeling the dynamics of open quantum systems \cite{tanimura89}. While HEOM has previously been applied to molecular bond rupture at surfaces \cite{yaling23, yaling21,erpenbeck20}, nonequilibrium chemical reactions \cite{yaling22_reaction}, desorption from surfaces \cite{erpenbeck19}, and the scattering of a single atom from a surface \cite{shi23}, in this work, we present the first application of HEOM to the scattering of a molecule with multiple nuclear degrees of freedom from a metal surface. After establishing the general theoretical framework, we exemplify the approach via application to the longstanding problem of NO scattering from Au(111), where previous quantum calculations have relied on an approximate description of the coupling to the surface electrons \cite{monturet10}. We observe multi-quantum relaxation of the bond vibrational state due to coupling to EHPs in the surface, in accordance with experiment. In doing so, we demonstrate that the HEOM approach not only represents a significant step forward in the modeling of nonadiabatic scattering of molecules from metal surfaces, but that it can be further applied as a benchmark both for models and other approximate theoretical methods.

\section{Theory}
\label{theory}

\noindent Throughout this work, we use natural units such that $\hbar = e = m_e = 1$. We employ a Newns-Anderson Hamiltonian, which is the standard model for dynamics at metal surfaces \cite{newns69,anderson61}. The Hamiltonian is partitioned according to 
\begin{equation}
H = H_\text{mol} + H_\text{surf} + H_\text{coup}, 
\end{equation}
where $H_\text{mol}$ corresponds to the scattering molecule, $H_\text{surf}$ to the surface, and $H_\text{coup}$ to the coupling between the molecule and surface. The molecular Hamiltonian is given by
\begin{equation}
H_\text{mol} = \sum_\kappa \frac{p_\kappa^2}{2m_\kappa} + U_0(\mathbf{x}) + \sum_{ij} h_{ij}(\mathbf{x})d_i^\dag d_j,
\label{general ham}
\end{equation}
where $\mathbf{x}$ is the vector of nuclear coordinate operators, whose elements are indexed by $\kappa$, with reduced mass $m_\kappa$, and $p_\kappa$ is the set of corresponding conjugate momenta. The vibrational potential energy surface for the neutral molecule is given by $U_0(\mathbf{x})$. The $d_i^\dag /d_i$ are the creation/annihilation operators for the electronic state $i$ on the molecule, where the energies and interstate coupling strengths are contained within $h_{ij}$. The electrons in the metal are modeled as a noninteracting reservoir of electrons, such that
\begin{equation}
H_\text{surf} = \sum_k \epsilon_k c^\dag_k c_k,
\end{equation}
where the $c^\dag_k/c_k$ denote the creation/annihilation operators for the state $k$ with energy $\epsilon_k$. The surface is held at a local equilibrium, such that the occupation function for electrons and holes is
\begin{equation}
f^{\sigma}(\epsilon) = \left(1 + e^{\sigma(\epsilon - \mu) / k_B T}\right)^{-1},
\end{equation}
where $\sigma \in \{+,-\}$, with $\sigma=+$ for electrons and $\sigma=-$ for holes, while $T$ is the temperature and $\mu$ is the chemical potential. The electronic coupling between the molecule and the surface is given by
\begin{equation}
H_\text{coup} = \sum_{ki} V_{ki} (\mathbf{x}) \left( d_i^\dag c_k + c^\dag_k d_i \right),
\end{equation}
with coordinate-dependent coupling elements $V_{ki} (\mathbf x)$. 

After integrating out the surface degrees of freedom, the influence of the molecule-surface coupling on the dynamics of the molecule arises via the two-time bath correlation functions,
\begin{equation}
C^\sigma_{ij} (t) = \frac{1}{2\pi}\int^\infty_{-\infty}e^{i\sigma\epsilon t}\Gamma_{ij} (\epsilon) f^{\sigma}(\epsilon)d\epsilon ,
\label{BCF}
\end{equation}
where $\Gamma_{ij} (\epsilon)$ is the spectral density function, defined as 
\begin{equation}
\Gamma_{ij} (\epsilon, \mathbf x) = 2\pi \sum_k  V^*_{ki}(\mathbf x)V_{kj}(\mathbf x)\delta(\epsilon - \epsilon_k).
\label{gam_orig}
\end{equation}
The energy dependence of the spectral density is assumed to take the form of a Lorentzian, such that 
\begin{equation}
\Gamma_{ij} (\epsilon, \mathbf x) =  2\pi V^*_{i}(\mathbf x)V_{j}(\mathbf x)  \frac{W^2}{(\epsilon - \mu )^2 + W^2},
\label{gam}
\end{equation}
where $W$ is the bandwidth and the strength of the electronic coupling to the surface is described by $V_i$. A fundamental step in the HEOM approach is to decompose the bath correlation functions in Eq.(\ref{BCF}) as a power series of exponential functions. For a finite temperature, it can be accurately approximated with only a small, finite number of poles, $P$, such that
\begin{equation}
C_{ij}^\sigma (t) \approx  V^*_{i}(\mathbf x)V_{j}(\mathbf x)\sum_{p=1}^{P}  \eta_{p\sigma} e^{-\gamma_{p\sigma} t}.
\label{pole decomp}
\end{equation}
In this work, we employ the barycentric sum-over-pole decomposition scheme to calculate $\eta_{p\sigma}$ and $\gamma_{p\sigma}$, as detailed in Refs.~\citen{barycentric23,takahashi24}. The decomposition in Eq.(\ref{pole decomp}) can be interpreted as a mapping of the continuum of electronic states in the metal onto a finite set of $K = 2N_e P$ effective virtual fermionic bath states, where $N_e$ is the number of electronic degrees of freedom of the molecule. 

The dynamics of the system can then be expressed as a hierarchy of equations of motion \cite{yaling22,tanimura20,tanimura89,jin07,jin08,schinabeck18,hartle15,zheng09,yan14,wenderoth16,ye16}, 

\begin{widetext}
\begin{multline}
\frac{d\rho^{(\mathbf n)}(t)}{dt} = 
\
-i\left[H_\text{mol}, \rho ^{(\mathbf n)}(t) \right]
\
- \sum_{k=1}^K n_k \gamma_{p_k \sigma_k }\rho^{(\mathbf n)}(t)
\\
+i\sum_{l=1}^K\sum_{k=1}^K \delta_{\bar\sigma_k, \bar\sigma_l} \delta_{p_k, p_l} (-1)^{\sum_{j<k} n_j} \sqrt{1-n_k} V_{i_k} \left(d_{i_k}^{\bar\sigma_k} \rho^{(\mathbf{n+1})_l}(t) + (-1)^{|\mathbf{n}|+1}\rho^{(\mathbf{n+1})_l}(t)d^{\bar\sigma_k}_{i_k}\right)
\
\\
\
+i\sum_{k=1}^K (-1)^{\sum_{j<k} n_j} \sqrt{n_k} V_{i_k} \left(\eta_{p_k \sigma_k} d_{i_k}^{\sigma_k} \rho^{(\mathbf{n-1})_k}(t) - (-1)^{|\mathbf{n}|-1}\rho^{(\mathbf{n-1})_k}(t)\eta^*_{p_k \bar\sigma_k}d^{\sigma_k}_{i_k}\right).
\label{HEOM}
\end{multline}
\end{widetext}
Here, $\mathbf{n} = [n_1,n_2,...,n_K]$, where $n_m$ is the occupation of the $m^\text{th}$ virtual bath state, which can be either $0$ or $1$. The norm is $\rvert n \rvert = \sum_{k=1}^K n_k$ and $\bar \sigma = -\sigma$. The creation and annihilation operators have also been rewritten in this notation, such that $d^{+}_{i_k} = d^{\dag}_{i_k}$ and $d^{-}_{i_k} = d^{}_{i_k}$. The reduced density matrix for the scattering molecule is obtained when all virtual bath states are unoccupied, such that $\rho_\text{mol} = \rho^{([0,0,...,0])}$. Each subsequent configuration of bath occupations, $\mathbf{n}$, then results in an additional auxiliary density operator (ADO), $\rho^{(\mathbf{n})}$, which collectively contain information about the influence of the bath on the dynamics of the molecule. The hierarchy is formed via the coupling of each ADO, $\rho^{(\mathbf{n})}$, up/down to ADO's of a higher/lower tier, $\rho^{(\mathbf{n\pm1}_k)}$, where
\begin{equation}
\mathbf{n} \pm \mathbf{1}_k = [n_1, n_2 ,..., 1-n_k ,..., n_K].
\end{equation}

The direct propagation of Eq.(\ref{HEOM}) quickly becomes computationally infeasible for systems with a large dimensionality of the Hilbert space, limiting its application to scattering systems consisting either of only a single nuclear degree of freedom, or restrictively small basis sets. Here, we overcome this limitation and enable the treatment of realistic molecular scattering systems consisting of multiple nuclear degrees of freedom, by reformulating the HEOM into a Schr\"odinger-like equation of motion for an extended wavefunction in twin space, which then facilitates the use of a matrix product state (MPS) representation of the state of the system \cite{shi18,borrelli19,takahashi24}. In doing so, the dimensionality of the molecular density matrix and ADOs scales polynomially with respect to the size of the vibrational basis set rather than exponentially, which is especially critical for the accurate treatment of vibrational, rotational, and translational degrees of freedom of a molecule where a large basis set is often needed. 

In the reformulation \cite{yaling22}, the ADO's are recast as tensors in the twin space according to
\begin{equation}
\rvert \rho^\mathbf{n}(t)\rangle = \sum_{s_1\tilde s_1 ... s_D \tilde s_D} C^\mathbf{n}_{s_1\tilde s_1 ... s_D \tilde s_D} \rvert s_1 \rangle \otimes \rvert \tilde s_1 \rangle \otimes ... \otimes \rvert s_D \rangle \otimes \rvert \tilde s_D \rangle,
\end{equation}
where $D$ is the number of degrees of freedom of the molecule. Subsequently, each operator for the molecule in Eq.\eqref{general ham} adopts two twin-space counterparts defined according to 

\begin{align}
\hat{d}^\pm_i \rvert\rho\rangle &= d^\pm_i\otimes\mathds{1}_{e_i}\rvert\rho\rangle \equiv d^\pm_i \rho,
\\
\tilde{d}^\pm_i \rvert\rho\rangle &= \mathds{1}_{e_i} \otimes d^\mp_i\rvert\rho\rangle \equiv  \rho d^\mp_i,
\\
\hat{x}_\kappa \rvert\rho\rangle &=x_\kappa \otimes \mathds{1}_{x_\kappa} \rvert\rho\rangle \equiv  x_\kappa\rho,
\\
\tilde{x}_\kappa \rvert\rho\rangle &= \mathds{1}_{x_\kappa} \otimes x_\kappa\rvert\rho\rangle \equiv  \rho x_\kappa.
\end{align}
The operators with a hat act on the physical
degrees of freedom, while those with a tilde act on ancilla degrees of freedom. One can additionally define a set of ad-hoc creation/annihilation operators which act on the Fock state for the virtual bath states, $\rvert \mathbf{n}\rangle = \rvert n_1...n_K \rangle$. These, along with an additional parity operator, $I^>$, are defined as
\begin{align}
c_k^{\lessgtr,+}\rvert\mathbf{n}\rangle &= (-1)^{\sum_{j\lessgtr k}} \sqrt{1-n_k}\rvert \mathbf{n+1}_k\rangle,
\\
c_k^{\lessgtr,-}\rvert\mathbf{n}\rangle &= (-1)^{\sum_{j\lessgtr k}} \sqrt{n_k}\rvert \mathbf{n-1}_k\rangle,
\\
I^>\rvert\mathbf{n}\rangle &= (-1)^{\sum_{j= 1}^K n_j} \rvert \mathbf{n}\rangle.
\end{align}
The contribution of all ADOs are then contained within an extended pure state wavefunction for the molecule and surface, given by
\begin{equation}
\rvert \Psi (t) \rangle = \sum_{\substack{n_1,...,n_K \\ s_1,..,\tilde{s}_D}} C^{[n_1, n_2, ..., n_K]}_{s_1\tilde s_1 ... s_D \tilde s_D} \rvert n_1  n_2 ... n_K \rangle \rvert s_1 \tilde s_1 ... s_D   \tilde s_D \rangle.
\end{equation}
It is defined such that any ADO can be readily extracted via a projection onto the relevant Fock state for the bath,
\begin{equation}
\rvert \rho^\mathbf{n}(t) \rangle = \langle \mathbf{n} \rvert \Psi(t) \rangle.
\end{equation}
The time evolution of the extended wavefunction follows
\begin{equation}
i\frac{d\rvert\Psi\rangle}{dt} = \mathbb{H}\rvert \Psi \rangle,
\label{schrodinger}
\end{equation}
where the super-Hamiltonian, $\mathbb{H}$, is 

\begin{align}
\mathbb{H} =& H_\text{mol} - \tilde{H}_\text{mol}
\
-i\sum_{k=1}^K \gamma_{p_k\sigma_k} c_k^{<,+}c_k^{<,-} \nonumber
\
\\
\
&-\sum_{l=1}^K\sum_{k=1}^K \delta_{\bar\sigma_k, \bar\sigma_l} \delta_{p_k, p_l} V_{i_k} \left(c_l^{<,-}\hat d_{i_k}^{\bar \sigma_k} - c_l^{>,-}\tilde d_{i_k}^{\bar \sigma_k}\right) \nonumber
\
\\
\
&-\sum_{k=1}^K V_{i_k} \left(\eta_{p_k\sigma_k}c_k^{<,+}\hat d_{i_k}^{\sigma_k} - \eta^*_{ p_k\bar{\sigma}_k}c_k^{>,+}\tilde d_{i_k}^{\sigma_k}\right).
\end{align}
The form of Eq.(\ref{schrodinger}) is conducive to the use of tensor network schemes such as the matrix product state (MPS) formalism, which is employed here. In doing so, the high-rank coefficient tensor, $C^{[n_1, n_2, ..., n_K]}_{s_1\tilde s_1 ... s_D \tilde s_D}$, is decomposed as a product of low rank matrices

\begin{align}
C^{[n_1, n_2, ..., n_K]}_{s_1\tilde s_1 ... s_D \tilde s_D} = \sum_{r_0, r_1, \cdots, r_{K+2D}} & A^{[1]}(r_0, r_1, n_1)A^{[2]}(r_1, r_2, n_2)\cdots A^{[K]}(r_{K-1}, r_K, n_K)\nonumber
\\
&\times A^{[K+1]}(r_{K}, r_{K+1}, s_1)\cdots A^{[K+2D]}(r_{K+2D-1}, r_{K+2D},\tilde s_D),
\label{MPS}
\end{align}
where $A^{[i]}$ is a rank-3 tensor with two virtual indices ($r_{i-1}, r_i$) and one physical index. To enforce that the decomposition for a given element of the tensor of coefficients evaluates to a scalar, an open boundary condition is applied where $r_0 = r_{K+2D} = 1$. For an arbitrary state, the decomposition in Eq.(\ref{MPS}) is formally exact in the limit of infinite bond dimension, $r_i$. In practice, a truncation to a finite bond dimension is applied. The accuracy of the MPS representation of the state can then be tuned by manipulating the bond dimensions in the MPS.
The super-Hamiltonian is analogously decomposed as a matrix product operator,
\begin{align}
\mathbb{H} = \sum_{r_0, r_1, \cdots, r_{K+2D}} & X^{[1]}(r_0, r_1, n_1,n_1')\cdots X^{[K]}(r_{K-1}, r_K, n_K,n_K') \nonumber
\\
& \times X^{[K+1]}(r_{K-1}, r_K, s_1,s_1')\cdots X^{[K+2D]}(r_{K+2D-1}, r_{K+2D}, \tilde s_D, \tilde s_D').
\label{MPO}
\end{align}

For the propagation of the MPS in Eq.\eqref{schrodinger}, we employ the time-dependent alternating minimal energy (tAMEn) solver proposed by Dolgov \cite{dolgov19} and applied to HEOM by Borrelli et al. \cite{borrelli21,taka24}, in which the MPS bond dimensions and time-step size are adaptively altered throughout the propagation according to a specified error tolerance. The results presented here are converged with respect to this error tolerance.  

\section{Results}
\label{results}

\subsection{Model}
\label{model}

\noindent The utility of the introduced method is now demonstrated via its application to the modelling of NO/Au scattering. The model employed here is taken from Ref.~\citen{gardner23} and is parameterized for the scenario when the NO molecule is incident on the surface with the N atom facing down towards the surface \cite{meng22}. A reduced description of the NO molecule is considered in which only two nuclear degrees of freedom are retained; the center of mass distance from the surface, $z$, and the NO bond stretching, $r$. We additionally consider the molecule to have only a single electronic level that participates in the dynamics, which can be either occupied or unoccupied. The diabatic potential energy surfaces that describe the molecule are given by
\begin{equation}
U_0 (r,z) = U_M(r-r_0; D_0, a_0) + \exp[-b_0 (z - z_0)] + c_0,
\label{U0}
\end{equation}
\begin{equation}
U_1 (r,z) = U_M(r-r_1; D_1, a_1) + U_M(z-z_1; D_2, a_2) + c_1,
\label{U1}
\end{equation}
where $U_M$ is the Morse potential, defined as
\begin{equation}
U_M(x;D,a) = D[\exp(-2ax) - 2\exp(-ax)].
\end{equation}
The two diabatic states represent the neutral state, $U_0$, and a negatively charged anionic state, $U_1$, of the molecule. The parameters in Eq.\eqref{U0} and \eqref{U1} were determined by parametrization against first principles data based on constrained density functional theory \cite{meng22}, where the number of electrons in the molecule was constrained to a fixed value.

\begin{figure}
\includegraphics[scale=0.67]{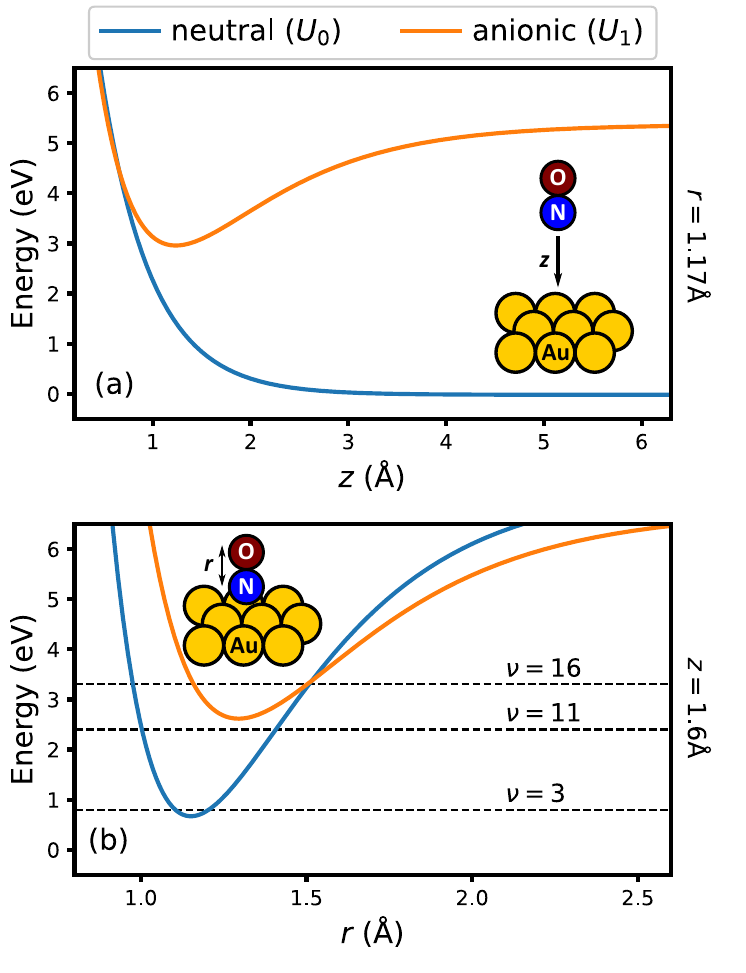}
\caption{One-dimensional cuts of the neutral and charged potential energy surfaces. (a) As a function of $z$ for $r = 1.17$ \AA. (b) As a function of $r$ for $z = 1.6$ \AA, where the energies of the $\nu\in\{3,11,16\}$ states are also shown.}
\label{pot_cross}
\end{figure}

The molecular Hamiltonian is then given by
\begin{equation}
H_\text{mol} = \frac{p_z^2}{2m_z} + \frac{p_r^2}{2m_r} + U_0(r,z) + [U_1(r,z) - U_0(r,z)] d^\dag d.
\end{equation}
The masses $m_z$ and $m_r$ correspond to the total mass of the NO molecule and the reduced mass for the bond, respectively. Example one-dimensional cuts along each axis of the potentials are shown in Fig.~\ref{pot_cross} and these will act as a helpful guide for the discussions in the following sections. 
The electronic coupling strength is taken to be dependent only on $z$ with two parameters that have been chosen to best approximate the ground state adiabatic potential energy surface for the NO/Au(111) system. It takes the form
\begin{equation}
V(z) = V[1-\tanh(z/\tilde{a})].
\end{equation}
The temperature of the surface is set to $300$K. The full set of model parameters can be found in Section 1 of the supplementary material or in Ref.~\citen{gardner23}.

The initial wavepacket for the molecule is described by a product state, in which the bond stretch mode is initialised in an eigenstate of $U_0$ in the limit $z\rightarrow\infty$. The eigenstates are denoted by $\nu$ while the initial state is $\nu_\text{ini}$. The $z$-component of the initial wavepacket is taken to be a Gaussian wavepacket incident on the surface with an average kinetic energy, $\text{KE}_\text{ini} = p_\text{ini}^2/2m_z$, as given by 
\begin{equation}
\Phi(z) = \frac{1}{(2\pi \sigma^2)^\frac{1}{4}}\exp\big[-\frac{(z-z_\text{ini})^2}{4\sigma^2}+i\frac{p_\text{ini}}{\hbar}(z-z_\text{ini})\Big].
\label{z_state}
\end{equation}
The parameters in Eq.\eqref{z_state}, along with all other associated parameters and details of the computation, are provided in Section 2 of the supplementary material.

\subsection{Energy transfer dynamics of molecular scattering}
\label{results}

\begin{figure}
\includegraphics[scale=0.6]{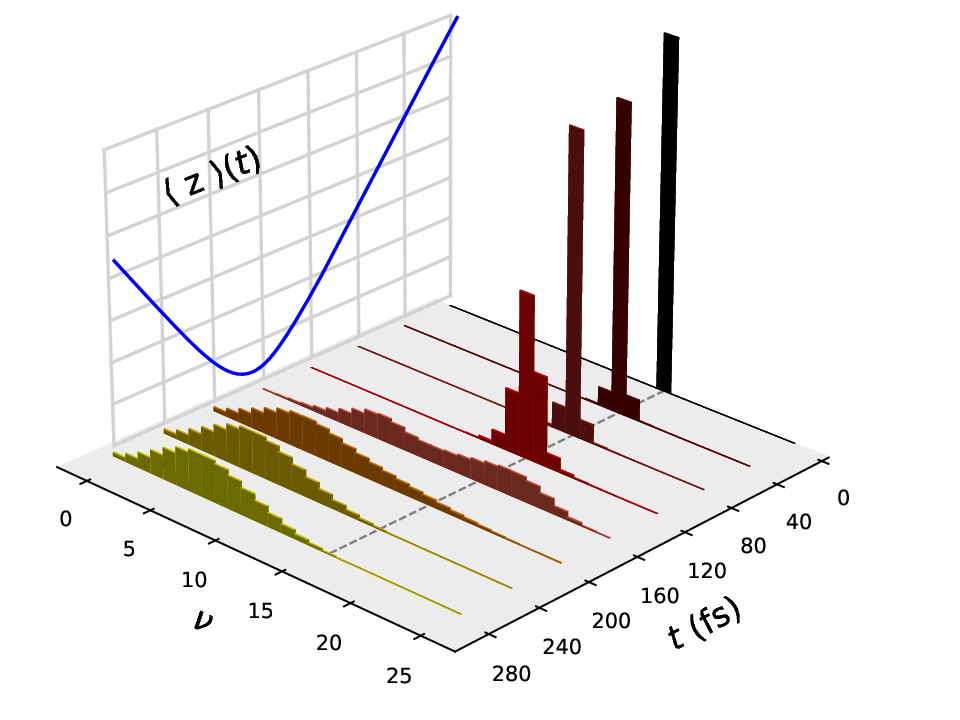}
\caption{Time dependence of the vibrational state populations, $P_\nu(t)$, with $\nu_\text{ini} = 16$ and $\text{KE}_\text{ini} = 1.0$ eV. The background shows the time dependence of $\langle z \rangle$. }
\label{bond_time_dep}
\end{figure}

\noindent Here, we present exploratory results of HEOM simulations for the scattering process. As a first demonstration, we calculate the time dependent populations of vibrational states of the bond stretching mode according to 
\begin{equation}
P_\nu (t) = \text{Tr}\{\rvert \chi_\nu \rangle \langle \chi_\nu \rvert \rho_\text{mol}(t)\},
\end{equation}
where $P_\nu$ is the population of state $\nu$, obtained via a projection onto the corresponding basis state $\rvert \chi_\nu \rangle$. Fig.~\ref{bond_time_dep} shows the time dependence of the population of bond vibrational states during the scattering process when initialized in the state $\nu_\text{ini} = 16$ with an initial translational kinetic energy of $\text{KE}_\text{ini} = 1.0\,\text{eV}$. In addition, the average distance from the surface $\langle z \rangle (t)$ is depicted. On approach towards the surface, the bond experiences both relaxation and excitation from the initial state due to energy exchange with electrons in the surface. At the point of closest approach, when $\langle z \rangle$ is at the minimum, the vibrational distribution is bimodal, where the occupation of higher vibrational states is also correlated with a higher electronic occupation of the molecule (see Fig. S1 in the supplementary material). Upon scattering away from the surface, the highly excited vibrational bond states relax to lower energy states and the state of the bond converges to the final distribution.

\begin{figure}
\hspace{-0.5cm}
\;\;\;\;\;\includegraphics[scale=0.77]{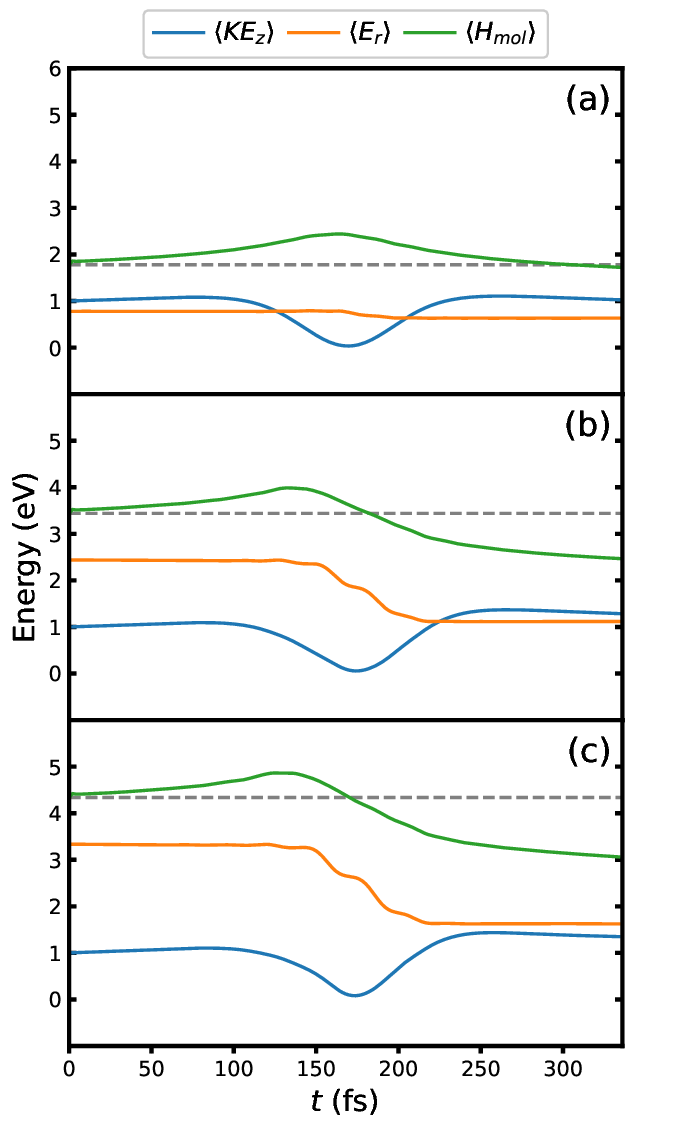}
\caption{Time dependence of the mean molecular energy along with its components, the mean translational kinetic energy and the energy stored in the molecular bond. $\text{KE}_\text{ini} = 1.0$ eV with (a) $\nu_\text{ini} = 3$, (b) $\nu_\text{ini} = 11$, and (c) $\nu_\text{ini} = 16$. The dashed line serves as a reference for the initial $\langle H_\text{mol} \rangle$.}
\label{energies_time}
\end{figure}

It is insightful to analyse the time dependent components of the energy in the system to better understand the dynamical processes at play. In Fig.~\ref{energies_time}, the time dependence of the molecular energy, $\langle H_\text{mol} \rangle$, along with the corresponding translational kinetic energy, $\langle\text{KE}_\text{z}\rangle$, and bond energy, $\langle E_r\rangle$ are shown for three different initial vibrational states where $\text{KE}_\text{ini} = 1.0~\text{eV}$. The bond energy is defined as
\begin{equation}
\langle E_r\rangle = \langle \text{KE}_r + U_0(r,z\rightarrow\infty)\rangle.
\end{equation} 
In all cases, the loss of energy of the molecule arises almost entirely via the relaxation of the vibrational state of the bond. This energy is primarily lost to the surface in the form of electron-hole pair excitations. However, some of the energy is instead transferred to the translational kinetic energy of the molecule upon scattering. This is evidence for EHP-mediated weak translation-vibration coupling, which has been previously observed in NO scattering experiments \cite{golibrzuch13}. By virtue of considering only a 2D model, energy transfer to other nuclear degrees of freedom such as the molecule's rotational motion is not considered \cite{golibrzuch13,shenvi09,serwatka20}.

The loss of energy of the molecule to the surface electrons, calculated as the change in $\langle H_\text{mol} \rangle$ over the scattering event, can be used to quantify the prevalence of nonadiabatic effects contributing to the dynamics, since energy can only be lost to the surface via nonadiabatic EHP excitation effects. In the $\nu_\text{ini}=3$ case, the molecule loses $0.1~\text{eV}$ to the surface electrons throughout the scattering process, whereas a higher initial vibrational state yields a much larger energy loss; $1.0~\text{eV}$ for $\nu_\text{ini}=11$ and $1.3~\text{eV}$ for $\nu_\text{ini}=16$. This can be rationalized via reference to Fig.~\ref{pot_cross}(b), where the energy of the $\nu = 16$ state is near to the energy of intersection between the neutral and charged potentials when the molecule is close to the surface. As a result, energy exchange processes between the molecule and the electrons in the surface are more accessible in comparison to when the molecule is initialized in a less excited state. When initialised in the $\nu_\text{ini}=3$ state, the molecule interacts comparatively weakly with the electronic states in the surface at all times throughout the scattering event. 

The energy lost by the molecule is also strongly dependent on its incoming kinetic energy. As demonstrated in Tbl.~\ref{energy_table} for a molecule initialized in the $\nu_\text{ini}=11$ state, the energy loss of the molecule to the EHPs is small for $\text{KE}_\text{ini}=0.2~\text{eV}$, but becomes significantly larger as $\text{KE}_\text{ini}$ increases. This behaviour is clarified by Fig.~\ref{pot_cross}(a), in which a large kinetic energy is required to climb the potential barrier before being able to strongly interact with the electrons in the surface. Furthermore, the energy transfer from the bond degree of freedom to the translational motion of the molecule, as mediated by the electrons in the surface, becomes proportionally larger with increasing $\text{KE}_\text{ini}$.

\begin{table}[h]
\centering
\setlength{\tabcolsep}{0pt}
\renewcommand{\arraystretch}{1.5} 
\begin{tabularx}{\columnwidth}{@{} >{\centering\arraybackslash}X >{\centering\arraybackslash}X >{\centering\arraybackslash}X >{\centering\arraybackslash}X @{}}
\rowcolor{gray!20}
$\text{KE}_\text{ini} = $ & $0.2~\text{eV}$ & $0.5~\text{eV}$ & $1.0~\text{eV}$ \\ 
Energy lost from bond & $0.17~\text{eV}$ & $0.89~\text{eV}$ & $1.32~\text{eV}$ \\ 
\% lost to EHPs & $55.3$ & $77.3$ & $73.9$ \\ 
\% transferred to $\text{KE}_z$ & $9.6$ & $15.8$ & $21.5$ \\ 
\end{tabularx}
\caption{Energy loss pathways for $\nu_\text{ini} = 11$. The displayed percentages are in terms of the energy lost from the bond degree of freedom.}
\label{energy_table}
\end{table}

\begin{figure*}
\begin{minipage}{0.49\textwidth}
\includegraphics[scale=0.53]{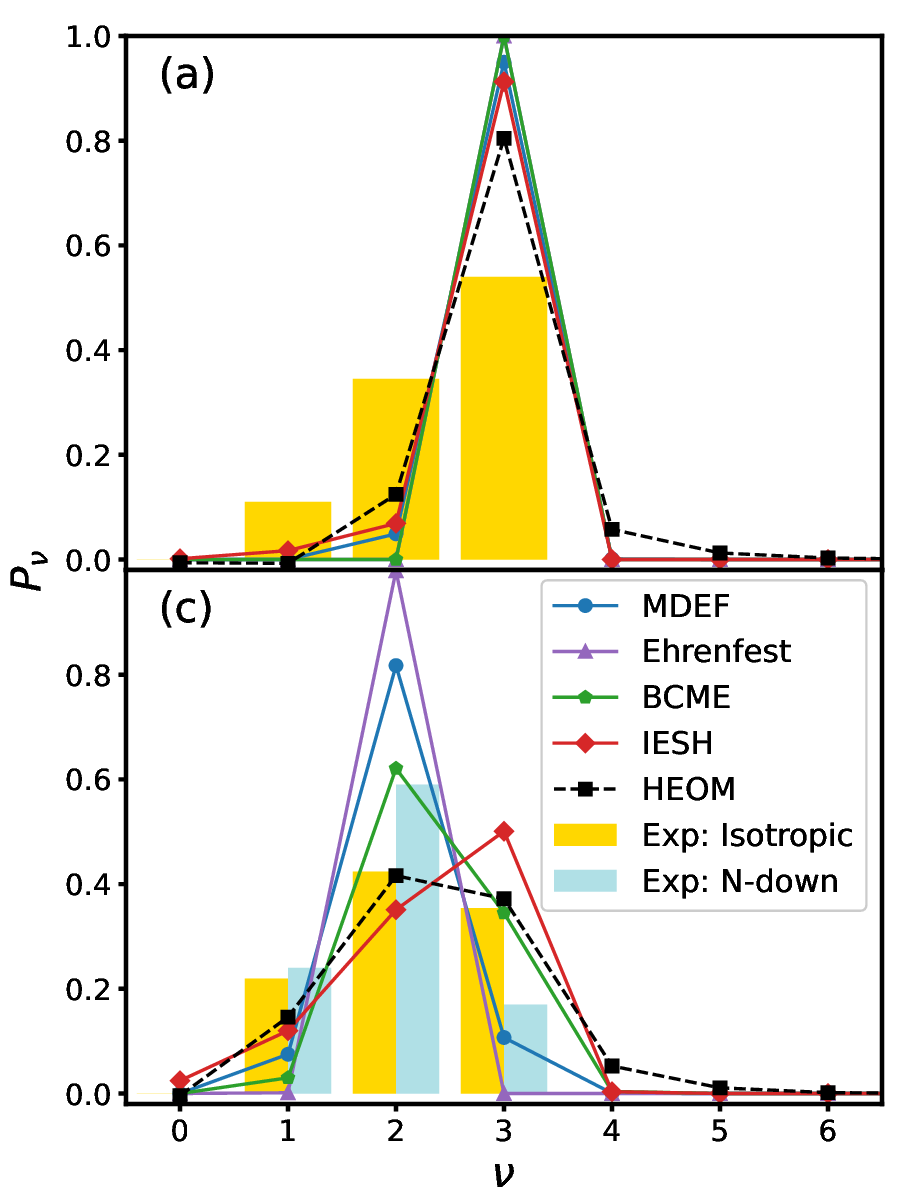}
\end{minipage}
\begin{minipage}{0.49\textwidth}
\includegraphics[scale=0.53]{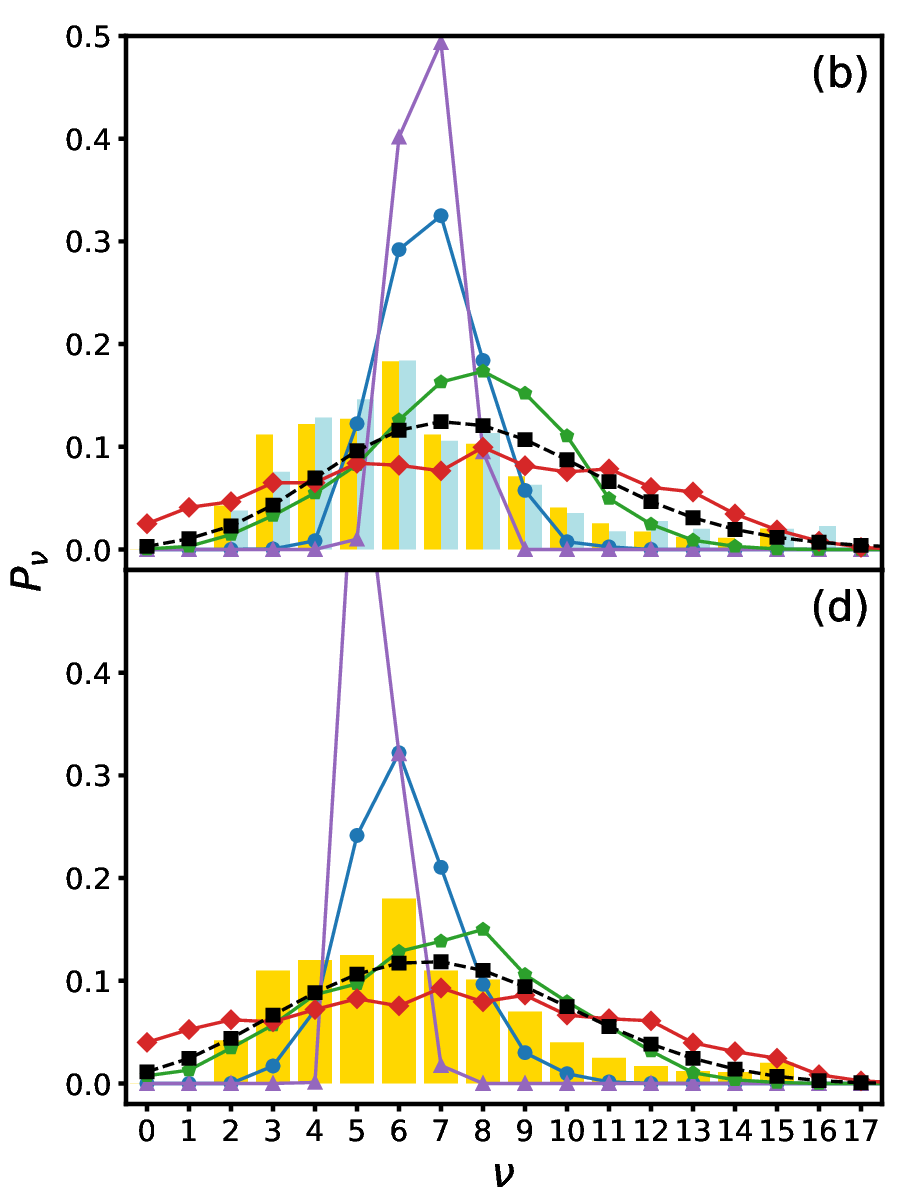}
\end{minipage}
\caption{Population of bond vibrational states after scattering. The left column corresponds to $\nu_\text{ini} = 3$ with (a) $\text{KE}_\text{ini} = 0.5$ eV and (c) $\text{KE}_\text{ini} = 1.0$ eV. The right column corresponds to $\nu_\text{ini} = 16$ with (b) $\text{KE}_\text{ini} = 0.5$ eV and (d) $\text{KE}_\text{ini} = 1.0$ eV. Colored lines show MQC calculations taken from Ref.~\protect\citen{gardner23}, where we include molecular dynamics with electronic friction (MDEF), Ehrenfest dynamics, broadened classical master equations (BCME), and independent electron surface hopping (IESH). The isotropic experimental data is taken from Ref.~\protect\citen{kruger15}. The N-down experimental data for $\nu_\text{ini}=3$ and $\text{KE}_\text{ini} = 0.95$ eV is taken from Ref.~\protect\citen{box21}, while the N-down experimental data for $\nu_\text{ini}=16$ and $\text{KE}_\text{ini} = 0.52$ eV is taken from Ref.~\protect\citen{bartels14}. All experimental data has been normalized.}
\label{vib_bench}
\end{figure*}

\begin{figure}
\hspace*{-1cm}
\includegraphics[scale=0.58]{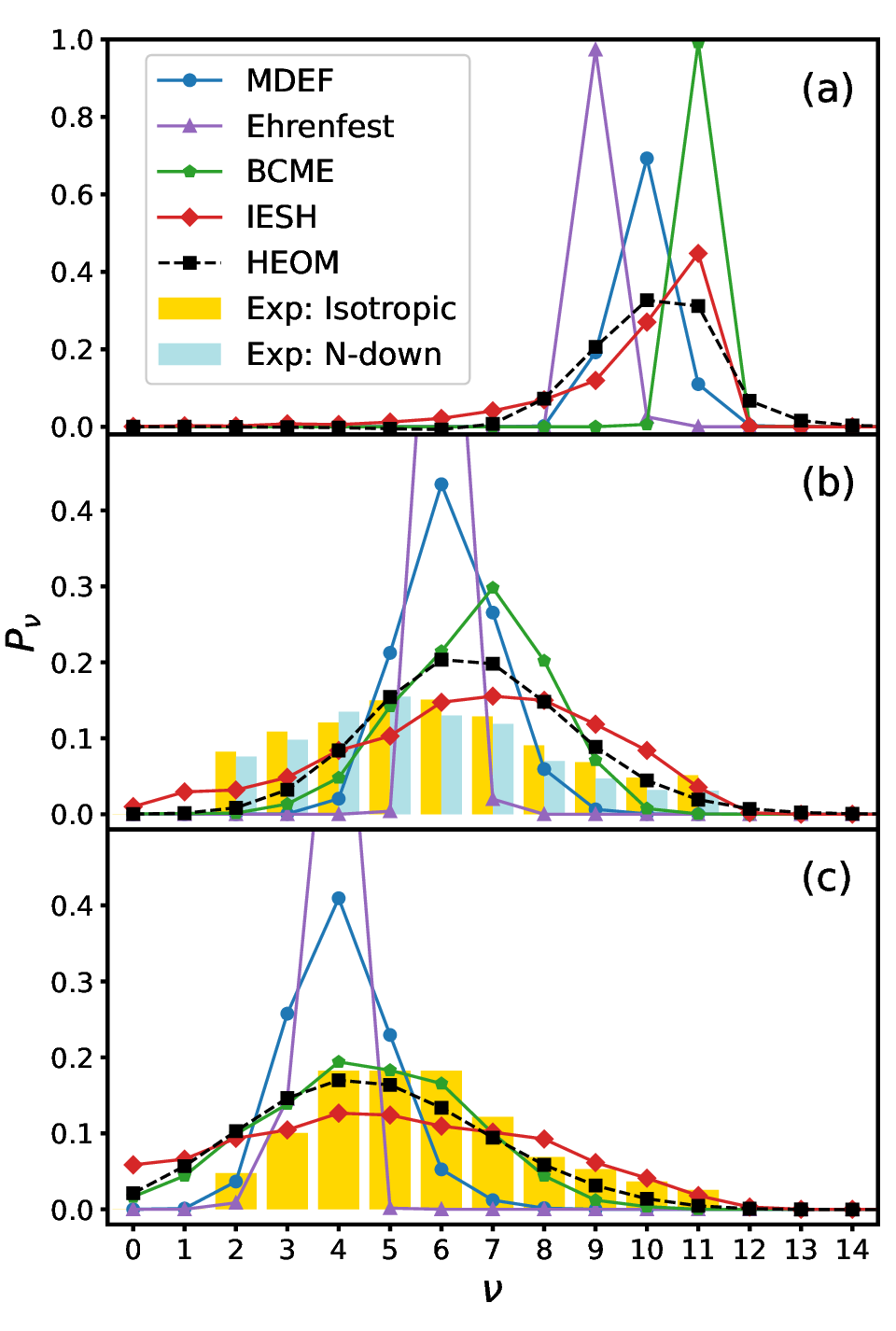}
\caption{Population of bond vibrational states after scattering for $\nu_\text{ini} = 11$ for (a) $\text{KE}_\text{ini} = 0.2$ eV (b) $\text{KE}_\text{ini} = 0.5$ eV (c) $\text{KE}_\text{ini} = 1.0$ eV. The isotropic experimental data is taken from Ref.~\protect\citen{kruger15}. The N-down experimental data for $\text{KE}_\text{ini} = 0.51$ eV is taken from Ref.~\protect\citen{bartels14}. All experimental data has been normalized.}
\label{vib_bench11}
\end{figure}

\subsection{Final vibrational state distributions}
\label{results_bench}

\noindent In accordance with the bulk of previous literature for the system under consideration, we analyse the final vibrational state distributions of the bond stretching after scattering. Beyond providing insight into the role of EHPs in the relaxation of the bond stretch mode state, we additionally highlight the value of the HEOM approach as a benchmark reference technique. To accomplish this, the HEOM method is here compared against a range of methods that treat the vibrational degrees of freedom of the NO molecule classically, for the same model system and the same physical parameters. Each MQC method considered incorporates the influence of nonadiabatic effects in the system via different assumptions. The comparison focuses on the final distribution of bond stretching state populations, which is also the main observable of interest. 

As a qualitative point of reference, we additionally include experimental data corresponding either to an isotropic initial orientation of the incoming molecule, or to cases where the incoming molecule has the N nucleus oriented down towards the surface; the latter serving as a more reliable reference for the model under consideration. By virtue of being numerically exact, comparison of the HEOM data with experimental results exposes the limitations associated with considering a reduced model of the scattering molecule and neglecting phonons in the surface. For example, consider Fig.~\ref{vib_bench}(c), which compares the final vibrational distributions after scattering when $\nu_\text{ini}=3$ and $\text{KE}_\text{ini} = 1.0$eV. The population of the $\nu=3$ vibrational quantum state is higher and the population of the $\nu=2$ state is lower in HEOM compared to the N-down experiment. This is likely due to the neglect of additional energy dissipation channels, such as energy transfer to phonons in the surface, or nuclear degrees of freedom beyond 2D such as the rotational motion of the molecule. This also emphasizes the importance of scrutinizing approximate dynamics methods of propagation with respect to numerically exact data for the same model, rather than directly comparing against experimental data. Having said that, we find that the experimental data is qualitatively reproduced by HEOM and the employed model in all parameter regimes tested here, particularly for higher vibrational initial states. This confirms that the model captures the essential aspects of nonadiabatic gas-surface scattering and provides a suitable platform to assess approximate MQC methods.

We now proceed with the benchmarking of MQC methods. In Fig.~\ref{vib_bench}(a), where  $\nu_\text{ini} = 3$ and $\text{KE}_\text{ini} = 0.5~\text{eV}$, all MQC methods differ from  the HEOM result, underpredicting both the relaxation and excitation from the initial vibrational state. In the case of $\text{KE}_\text{ini} = 1.0~\text{eV}$ as in Fig.~\ref{vib_bench}(c), the approaches based on the broadened classical master equation (BCME) and independent electron surface hopping (IESH) best approximate the HEOM data, while molecular dynamics with electronic friction (MDEF) and Ehrenfest dynamics approaches less accurately account for the spread of the vibrational distribution. Such an analysis is also true upon considering an initial vibrational state of $\nu_\text{ini} = 16$, as in Fig.~\ref{vib_bench}(b) for $\text{KE}_\text{ini} = 0.5~\text{eV}$, and Fig.~\ref{vib_bench}(d) for $\text{KE}_\text{ini} = 1.0~\text{eV}$. In both cases, BCME and IESH perform best in comparison to the exact HEOM data and each becomes more accurate for a larger initial kinetic energy of the molecule. Ehrenfest dynamics and MDEF perform comparatively poorly, each greatly underpredicting the spread of the final vibrational distribution despite also predicting a large amount of vibrational relaxation.

We further study previously unpublished data of MQC methods in Fig.~\ref{vib_bench11}(a)-(c), where the final vibrational distribution is shown for an initial vibrational state $\nu_\text{ini} = 11$ and for initial kinetic energies $\text{KE}_\text{ini} \in \{0.2~\text{eV}, 0.5~\text{eV}, 1.0~\text{eV}\}$. When $\text{KE}_\text{ini} = 0.2$ eV, the molecule experiences only a small amount of vibrational relaxation. This is because the molecule requires more energy to reach the crossing between the $U_0$ and $U_1$ states where the interaction with EHPs is strongest. This case is of particular interest as nonadiabatic effects remain prevalent while quantum nuclear effects become more prominent in the low kinetic energy regime, which should lead to a natural breakdown of the MQC methods. BCME is inaccurate in this regime, predicting no vibrational relaxation at all, while MDEF and Ehrenfest dynamics each predict some relaxation but largely quantitatively differ from the exact result. IESH fairs best out of the methods tested in this regime, but is still not a quantitatively accurate representation of the exact data. The BCME and IESH methods each converge towards the exact result upon increasing the initial kinetic energy of the molecule. Meanwhile, the Ehrenfest and MDEF methods prove to be unreliable in all kinetic energy regimes.

To summarize the comparison with the MQC methods, except for the failure of the BCME approach for a very low initial kinetic energy, the exact data is approximated to a reasonable degree by both the BCME and IESH approaches. However, BCME slightly underpredicts the spread of the vibrational distribution and IESH slightly overpredicts the spread of the vibrational final state distribution in most cases. In contrast, Ehrenfest dynamics and MDEF are consistently inaccurate, greatly underpredicting the spread of the final distribution. The results presented here demonstrate the inadequacy of a mean-field description of the electronic forces acting on the nuclear vibrations in the scattering molecule, which is central to Ehrenfest dynamics. MDEF, meanwhile, relies on an assumption of weak nonadiabaticity, and is thus unreliable in capturing the extent of the nonadiabatic EHP excitation effects which drive the vibrational relaxation of the molecule. BCME and IESH are not reliant on these assumptions and are thus better suited to accurately treat the vibrational relaxation in this scattering system.

They also appear to become more accurate for larger initial kinetic energies of the molecule where a MQC approximation should be more reliable. This is consistent with previous data, where the BCME approach has been shown to accurately reproduce numerically exact data in classical regimes \cite{DouBCME}. In the lower kinetic energy regime, the molecule remains longer at the surface and has more time to interact with phonons, while directional steering effects and rotational-vibrational coupling are more important. Additionally, the quantum nature of nuclei is more important at lower kinetic energies, meaning quantum nuclear effects and nonadiabatic effects both contribute strongly to the dynamics. The further development of MQC methods in this regime would be of great value to better understand ultrafast surface dynamics.

The discrepancy between the exact theoretical data and the experimental results, where such comparable experimental results exist, also highlights the need to explore further expansions to the model to capture the rotational-vibrational coupling of the molecule, the orientation of the molecule with respect to the surface and the phonons in the surface. This is especially true since it has been shown that a significant amount of the vibrational relaxation can be accounted for via an accurate representation of the high-dimensional adiabatic potential energy surface \cite{rongrong19}.

\section{Conclusion}
\label{Conclusions}

\noindent We have presented a theoretical approach to simulate the scattering of molecules from metal surfaces that treats the electronic coupling of the molecule to the surface electrons in a fully quantum mechanical and numerically exact manner via the HEOM approach. The approach is facilited by a matrix product state representation of the state of the system, enabling the treatment of vibrational basis sets that were previously prohibitively large. As a demonstrative example of the method, we have considered the scattering of NO from Au(111), where experiments have shown the dynamics to be strongly dependent on the nonadiabatic electronic coupling to the surface, resulting in multi-quantum relaxation in the bond vibrations due to the scattering. The HEOM method in conjunction with the model employed qualitatively reproduces the physics observed in experiments and has potential to serve as a crucial tool for validating the underlying scattering models in relation to experimental data. 

Additionally, we have used our HEOM approach to benchmark a range of mixed quantum-classical approaches, from which we are able to identify the broadened classical master equation and surface hopping methods as the most accurate in capturing the nonadiabatic vibrational relaxation of the bond mode. Ehrenfest dynamics and MDEF perform poorly in comparison. The general poor performance of all tested MQC methods in the low kinetic energy regime, coupled with the current void in available experimental data, presents an intriguing opportunity for new experimental investigations. Our hope is that the presence of a numerically exact benchmark in the form of HEOM motivates the further development of improved MQC methods that can simulate the dynamics of high dimensional reactive chemical dynamics at surfaces. The theoretical approach introduced and applied here has the potential to treat higher dimensional systems and can incorporate a phonon bath in the metal surface, whose significance we have identified through comparison with available experimental data. With this in mind, further application of the HEOM method developed here to molecular scattering problems will be the subject of future work.

\section*{Associated Content}
\subsection*{Supporting Information}
\noindent Further details of the theory, parameters of the model, details of the computation, and additional results.

\section*{Author Information}
\subsection*{Corresponding Authors}

\noindent\textbf{Riley J. Preston} − Institute of Physics, University of Freiburg, Hermann-Herder-Strasse 3, 79104 Freiburg, Germany; orcid.org/0009-0008-9858-4951; Email: riley.preston@physik.uni-freiburg.de

\section*{Acknowledgements}
\noindent We thank Andre Erpenbeck for helpful discussions. This work was supported by the Deutsche Forschungsgemeinschaft (DFG) within the framework of the Research Unit FOR5099 “Reducing complexity of nonequilibrium systems” and by UK Research and Innovation (UKRI) through a UKRI Future Leaders Fellowship [MR/X023109/1] and a UKRI frontier research grant [EP/X014088/1]. R.B. acknowledges the support of the project CANVAS, funded by the Italian Ministry of the Environment and the Energy Security, through the Research Fund for the Italian Electrical System (type-A call, published on G.U.R.I. n. 192 on 18-08-2022). S.L.R. and R.J.P thank the Alexander von Humboldt Foundation for the award of Research Fellowships. Y.K. thanks the Swiss National Science Foundation for the award of a research fellowship. Furthermore, the authors acknowledge the support by the state of Baden-Württemberg through bwHPC and the DFG through Grant No. INST 40/575-1 FUGG (JUSTUS 2 cluster).

\clearpage

\bibliography{bibli}

\end{document}